\documentstyle[aasms4,epsf]{article}
\tighten
\begin{document}
\title{ The remarkable starburst-driven outflow in NGC~2782}
%
\author {Shardha Jogee\altaffilmark{1}, Jeffrey D. P. Kenney\altaffilmark{1}
\& Beverly J. Smith\altaffilmark{2}}
\authoremail{jogee@astro.yale.edu, kenney@astro.yale.edu, beverly@ipac.caltech.edu}
\altaffiltext{1}{Yale University Astronomy Department, New Haven, CT 06520-8101}
\altaffiltext{2}{ IPAC/Caltech, Pasedena, CA 91125}

\abstract{
We show that  the starburst-driven outflow in the peculiar 
galaxy  NGC~2782 forms a well-defined  collimated  
bubble which has an extent of $\sim$ 1 kpc 
and a closed shell at its edge, as seen in H$\alpha$, [O~III], 
and 5 GHz radio continuum. The  shell coincides with the 
maximum in intensity and linewidths of [O~III] lines in a blueshifted 
emission nebula which was previously detected via 
optical spectroscopy by Boer et al. (1992). 
Such a remarkable outflow morphology has not been observed to date 
in any other starburst galaxy of comparable luminosity. 
The radio continuum map reveals a second bubble 
of similar size on the opposite side of the nucleus, forming 
a striking double-bubble outflow morphology. 
We argue from the morphology and short timescale 
($\sim$ $4 \times 10^6$ years) of the outflow that it 
is dynamically younger than freely-expanding outflows seen 
in other galaxies which  harbor circumnuclear starbursts 
of comparable  luminosity, e.g., M82. 
We suggest that the outflow in NGC~2782 is in the early 
stage where thermal instabilities have not yet completely 
ruptured the outflow bubble.
We present evidence that the outflow is driving 
warm and hot ionised gas, and possibly 
cold molecular gas, out of the central kpc of the galaxy. 
We estimate the  contribution of the hot, warm, and cold  phases of 
the ISM  to the energetics of the outflow.
This study is  based on our optical BVR, H$\alpha$, and [O~III]
observations from the WIYN telescope and OVRO CO interferometric
data, along with available 5 GHz radio continuum 
and ROSAT X-ray maps.
}

\section {Introduction}
Starburst-driven winds can  influence the formation 
and evolution of galaxies. Such winds 
can drive metal-enriched material 
out of the galactic plane, thus affecting the 
mass-metallicity relation between galaxies, the 
metallicity-radius relation within galaxies and 
the enrichment of intracluster medium (e.g, Dekel \&
Silk 1986; Heckman, Lehnert, \& Armus 1993, and 
references therein).  In addition, the properties of  young stellar 
components built by central starbursts depend on 
the rate at which starburst-driven winds blow 
the ISM out of the galaxy. It is therefore important to study how 
starburst-driven winds are triggered and evolve. 
A starburst-driven outflow can be associated with a 
large spectrum of emission e.g., radio continuum (RC), 
optical, UV, and X-rays. 
The relative intensity and distribution of the emission 
at different wavelengths give  critical clues about the 
evolutionary phase of the outflow and physical conditions in the ISM. 
Nuclear outflows have been observed in a large number of 
starburst and Seyfert galaxies (e.g., M82, 
NGC~253,  NGC~3628, NGC~3079, Arp 220, NGC~6240, NGC~4666, 
NGC~4945, NGC~1068, NGC~1320, 
NGC~4235) as discussed by many authors 
(see Heckman et al. 1993; 
Baum et al. 1993; Lehnert \& Heckman 1995; 
Colbert et al. 1996; Dahlem et al. 1997,  \& references therein). 
In this paper, we describe the starburst-driven outflow in the 
peculiar galaxy NGC~2782; 
the outflow has a remarkable morphology 
that has not been observed to date in any other starburst galaxy 
of comparable luminosity.
This study is based on  our  optical 
BVR, H$\alpha$, and [O~III] images
from the WIYN (Wisconsin Indiana Yale NOAO) telescope 
and 2$''$ resolution OVRO  (Owens Valley Radio Observatory) 
CO (J=1-$>$0) data, along with available 5 GHz RC
observations (Saikia et al. 1994), and an
archival ROSAT X-ray map.
In this paper, we focus on the starburst-driven
outflow and describe the optical observations in 
detail. In paper II  (Jogee, Kenney \& Smith 1997), we 
describe the observations and  full analysis of the CO data, 
and present evidence for a nuclear bar fuelling molecular 
gas into the central starburst.

\section {Previous work on the inner few kpc of NGC~2782}

NGC~2782  is a nearby (D=34 Mpc for H$_o$ of 75 km s$^{-1}$ Mpc$^{-1}$), 
peculiar galaxy with a pair of HI and optical tails 
(Sandage \& Bedke 1994; Smith 1991), and 
three ripples within the disk (Arp 1966; Smith 1994). 
The HI and optical properties 
can be modelled  with a recent interaction or 
merger of two disk galaxies of unequal mass (Smith 1994).
Previous H$\alpha$ maps  (Hodge \& Kennicutt 
1983; Smith 1994; Evans et al. 1996) showed an arc of emission along 
the inner ripple at r $\sim$ 25 $''$ (4.2 kpc), and an unresolved 
bright region of star formation in the inner 10$''$ (1.7 kpc) radius. 
This region harbors one of the most luminous 
circumnuclear starbursts among nearby  (D $<$ 40 Mpc) 
spirals (Devereux 1989), with a FIR luminosity 
($2 \times$  10$^{10} L_{\sun}$; Smith 1991) comparable to M82. 
The circumnuclear region  has optical 
emission spectra indicative of  HII regions (Sakka et al. 1973; 
Balzano 1983; Kinney et al. 1984), as well as an 
additional component of  highly excited, highly 
ionised gas (Kennicutt et al. 1989).
Boer, Shultz, and Keel (1992) used spatially resolved spectra to 
show that the high excitation gas lies in a blueshifted outflowing 
emission nebula. The 5 GHz RC map at 1$''$ resolution 
presented by Saikia et al. (1994) 
revealed an intriguing set of  peaks and bubbles, 
for which no explanation has yet been proposed. 
The detailed distribution and kinematics 
of molecular gas in the central region was previously 
unknown as NGC~2782 had been 
mapped only with single dish observations (e.g., Young 
et al. 1995)  and low resolution (6$''$) interferometry 
with the Nobeyama Millimeter Array (Ishizuki 1994). 
The latter study shows an elongated feature with two barely
resolved CO peaks. The higher (2$''$) resolution CO data 
we present in paper II (Jogee et al. 1997) resolves 
the elongated feature into a clumpy, double-lobed bar-like 
feature of radius $\sim$ 7.5$''$ (1.3 kpc).

\section {The remarkable outflow morphology revealed by new observations}

Two 5 min. exposures of NGC~2782 were taken in  H$\alpha$+[N II], 
[O~III], off-line continuum and Harris BVR  filters on 
the 3.5 m WIYN telescope at KPNO in December 1995 and March 1996 
A  2048 x 2048 S2KB CCD with a plate scale of 0.2$''$/pixel was used, 
giving a field of view of 6.8'x 6.8' (69 kpc x 69 kpc). 
Exposures were taken in on-line and off-line filters centered on 
6618 \AA\ and 6709  \AA\ for 
the redshifted H$\alpha$+[N II] 
$\lambda$$\lambda$ 6563 6583 \AA\ emission lines, and 
on 5055  \AA\  and 5176  \AA\ for the redshifted 
[O~III] $\lambda$$\lambda$ 4959 5007 \AA\ emission lines. 
The bandwidths of the four filters were 74, 71, 32, and 
52 \AA\ respectively. 
The average seeing in the images was 0.8$''$ except for the B and 
[O~III] images where it was $\sim$ 1.3$''$. 
We used the IRAF package 
to obtain continuum-free [O~III] and H$\alpha$+[N II] 
(thereafter referred to as H$\alpha$) images. 
The registration accuracy of the optical images is 
$\sim$ 1$''$.

The full field of view WIYN R image (Fig. 1, Plate L1) 
shows the two stellar tails, the optical disk, and the 
three ripples at radii of 25$''$, 45$''$, and 60$''$.
Except for these three ripples, the optical disk 
looks  relatively undisturbed  within a radius 
of 1$'$ (10 kpc). A magnified view of the inner disk region, as seen in 
the B-V image, is shown in Fig. 2a  (Plate L2).
It reveals two straight dust lanes which are offset from 
the nucleus and extend out to the inner ripple. 
The northern dust lane is redder and more prominent 
than the southern one. 
The dust lanes bracket the circumnuclear region of star formation 
which is shown in H$\alpha$ in Fig. 2b (Plate L2).
This H$\alpha$ map has a higher resolution (0.8$''$) than 
previously published  maps and resolves the 
region of star formation into 
(i) 
a centrally concentrated   H$\alpha$ peak, 
 (ii) 
\rm 
a clumpy, arc-like feature about $3 ''$ (0.5 kpc) 
north of the peak, extending east-west over 10$''$, and 
(iii) 
\rm 
an  H$\alpha$ bubble with a shell feature $\sim$ 
6$''$ S of the nucleus.

In order to unravel the interplay between the molecular gas, 
the starburst, and the nuclear outflow, we compare this 
H$\alpha$ image with the [O~III] image, the CO data presented 
in Paper II, and the 5 GHz RC maps from Saikia et al. (1994). 
This striking comparison is shown in Figures 3a-3d (Plate L3).
The RC map shows a compact central component which is 
surrounded by secondary peaks to the east and west, and  two extended 
bubbles to the north and south. 
The central RC, H$\alpha$, and [O~III] peaks are coincident 
within the uncertainties, and lie between the two CO lobes 
(Fig. 3a). 
The optical emission lines in the inner 3$''$ radius  
show narrow line widths (full width 
half maximum (FWHM) $\sim$ 300  km s$^{-1}$) 
and line ratios typical of HII regions (Boer et al. 1992). 
Thus, the central RC,  H$\alpha$, and [O~III] peaks are 
likely produced by massive star formation. 
The secondary eastern and western RC peaks lie within the 
CO bar (Fig. 3a) in a region which is 
associated with a large amount of molecular gas, shows strong
H$\alpha$ emission, and has  optical line ratios 
consistent with star formation (Boer et al. 1992). 
Therefore, the secondary RC peaks are also probably due 
to  massive stars.

However, the  northern and southern RC bubbles 
may have a different origin. 
They are not associated with much molecular gas (Fig. 3a) 
and are elongated  along the CO kinematic minor axis 
which has a position angle of $\sim$ 165 $\deg$ (Paper II). 
The northern RC bubble has a 
more fragmented morphology than the southern one and does not 
have optical counterparts. 
The southern RC bubble is associated with the 
H$\alpha$ bubble (Fig. 3c) and a bright  [O~III] bubble 
(Fig. 3d), and all three bubbles show 
a shell feature at $\sim$ 6$''$ (1 kpc) south. 
At this position, optical spectra 
(Boer et al. 1992) show that the [O~III] lines have 
a strong extranuclear peak, a large FWHM 
of $\sim$ 580 km s$^{-1}$, a blueshift 
of  120 km s$^{-1}$, 
and an  [O~III]/H$\beta$  ratio of $\sim$ 4.
The [O~III] image in Fig. 3d (Plate L3) also shows an 
extranuclear peak which lies in the shell and 
has an intensity about half that of the nuclear peak. 
We thus propose that the southern  H$\alpha$, [O~III], and RC  
bubbles, as well as the northern RC bubble, 
are part of a starburst-driven outflow 
emanating from the nucleus. 
The absence of an  optical counterpart to the northern 
RC bubble can be explained in terms of extinction 
by dust in the optical disk of NGC~2782. 
The relatively undisturbed and circular appearance of 
the disk (Fig. 1, Plate L1), as well as 
kinematic evidence presented in Paper II, suggests a 
low inclination for the disk.
Furthermore, the near side of the disk is probably to 
the north since the dust lanes
are redder and more prominent (Mihalas \& Binney 1981) 
in the northern side of the disk (see Fig. 2a, plate L2). 
We therefore conclude that the northern  
outflow bubble lies behind the disk and 
is probably obscured from view at optical wavelengths 
by dust in the disk.

\section {The evolutionary phase of the starburst-driven outflow}

The starburst-driven outflow in NGC~2782 has 
a well-defined  collimated bubble morphology 
with a closed shell at its outer edge, as seen 
in H$\alpha$, [O~III], and RC. 
Such an unusual outflow morphology has not been 
previously observed in any other starburst galaxy 
of comparable luminosity (see e.g., Heckman et al. 1993; 
Lehnert \& Heckman 1995, \&  references therein). 
In most outflows, the 
H$\alpha$ and  non-thermal RC form complex sets of 
filaments and loops (e.g., M82, NGC~253,  NGC~3628, NGC~4666; see Heckman 
et al. 1993; Seaquist \& Odegard 1991; Bland \& Tully 1988; 
Fabbiano et al. 1992; Dahlem et al. 1997, \&  references therein). 
The soft X-ray and RC emission often extends further out than 
the H$\alpha$.
The known outflow most similar to
NGC 2782 is that in the LINER galaxy NGC 3079.  This galaxy
has an H$\alpha$ bubble of  radius 13$''$ (1.1 kpc), 
with a shell feature at its outer edge. However, 
in contrast to NGC~2782, 
the RC emission in NGC~3079 forms two highly fragmented halos which 
extend twice as far as the H$\alpha$ bubble 
(Ford et al. 1986; Veilleux et al. 1994).

What accounts for the remarkable outflow morphology in NGC~2782? 
Theory predicts that in the early stages of a starburst-driven outflow, 
supernovae and winds from massive stars inject energy into 
the ISM and produce a bubble of very hot gas 
and thermalised ejecta (e.g., Chevalier \& Clegg 1985; 
Tomisaka \& Ikeuchi 1988; Heckman, Armus, \& Miley 1990). 
The bubble expands, sweeping up ambient gas into a thin, dense, 
optically-emitting shell. 
Non-thermal RC can  be produced when relativistic plasma 
which is driven along the pressure gradient of 
the ISM, undergoes synchroton 
losses and inverse Compton scattering (e.g, Heckman et al. 1993; 
Baum et al. 1993; Colbert et al. 1996). Once the bubble is several times 
the scale height of the ambient gas (blowout phase), 
the swept-up shell fragments through thermal instabilities 
and allows the hot gas to expand freely. 
For instance, the starburst galaxy  M82 is likely 
to be in this post-blowout, freely-expanding 
phase (e.g., Heckman et al. 1990), where the 
dense shell has ruptured, allowing the RC-emitting plasma 
and the hot X-ray-emitting gas to escape beyond the optical nebula. 
On the other hand, NGC~3079 might be in the blowout, 
partially-ruptured bubble phase (Veilleux et al. 1994).
We suggest that the southern starburst-driven outflow in  NGC~2782 
is dynamically younger than these  outflows; 
it is probably in  a pre-blowout  or early stages  of a 
blowout phase where thermal instabilities 
have not yet completely ruptured the optical bubble. 
The electron densities 
(n$_e$ = 600 $\pm$ 200 cm$^{-3}$ from the 
[S II]$\lambda$ 6716 \AA/[S II]$\lambda$ 6730 \AA\ 
line ratio; Boer et al. 1992) 
in the optical shell of NGC~2782
are higher than densities  observed in the outflow region of 
many galaxies including 
M82, NGC~253, NGC~3079, NGC~3628, NGC~4527, NGC~4536, NGC~4666, 
NGC~7541, etc (e.g., see Lehnert \& Heckman 1996).
This suggests that 
the shell feature in NGC~2782 contains  highly compressed ionised gas. 
The likely enhancement of the magnetic field strength in the 
compressed material could explain the strong RC emission in the shell.

Our results on the outflow morphology at H$\alpha$, [O~III], 
and RC, together with  the short outflow dynamical 
timescale ($\sim$ $ 4\times 10^6$ years) found 
from optical spectra by  Boer et al. (1992), suggest 
that the outflow is dynamically young.
This young dynamical phase could be attributed to several factors. 
The  circumnuclear starburst in NGC~2782 may be younger than 
other starbursts which have a comparable luminosity, but a 
biconical freely-expanding outflow, e.g., M82. 
The dynamical timescale of the 
outflow in NGC~2782 is less than in M82, where we computed the 
timescale to be $\sim$ 10$^7$ years, using the size 
and expansion velocity of the optical nebula (e.g., Heckman et al. 1990). 
Furthermore, molecular gas is present 
around the circumnuclear starburst of NGC~2782 (see Fig. 3a, Plate L3)
and  it can remove energy from the 
outflow and slow down its expansion, for instance, 
by increasing  the rate of radiative cooling (Heckman et al. 1993). 
In addition, the presence of clouds can reduce the density 
contrast between the dense compressed shell and the hot fluid 
it encloses, thereby delaying the onset of 
thermal instabilities which rupture the shell. 

Knowledge of the distribution of hot ionised gas in the outflow 
would help assess its dynamical phase.  In Fig. 4a, 
we give the H$\alpha$ map of a larger region of NGC~2782, 
and in Fig. 4b, we compare this map with an archival 
ROSAT HRI X-ray map which has a 
resolution of $\sim$ 12$''$.
The X-ray map shows a bright core surrounded by
fainter emission extending out to a radius of 20"
(3.5 kpc).  
Unfortunately, the low resolution of the X-ray
map does not resolve the outflow bubbles, and makes it
difficult to separate hot gas driven out by
the current outflow and that blown out by previous
episodes of star formation.  Such past episodes are not
unlikely.  The H$\alpha$  image shows 
diffuse H$\alpha$ arcs, loops, and filaments 
around the central region 
of star formation (Fig. 4a, Plate L4), and this morphology 
is highly suggestive of gas blown out in energetic episodes 
associated with previous starburst-driven outflows or 
with the interaction. 
The timescales involved also leave open the possibility 
of recurrent starbursts; the current outflow has a dynamical 
timescale of only $\sim$ $ 4 \times 10^{6} $ years 
while the time of closest approach for the interaction is 
$\sim$  $ 2 \times 10^{8} $ years (Smith 1994). 
Although the X-ray map has a low resolution, it nonetheless 
suggests an interesting possibility. 
The X-ray emission is more elongated to 
the north than to the south, and the northern RC 
bubble is also more extended and fragmented than 
the southern RC bubble (Fig. 3a, Plate L3); together, 
these facts suggest that the northern part of
the outflow may have already broken out and
allowed the hot gas to escape. It is unlikely 
that the more extended northern X-ray emission is an 
instrumental artifact because a similar extension is not 
seen in a point-like source near NGC~2782 in the ROSAT 
map. Furthermore, it is striking that the X-ray emission appears 
to extend out to the northwestern H$\alpha$  arc whose narrow, 
bright appearance is suggestive of highly 
shocked gas (see Fig. 4b, Plate L4). 

\section {Cold, warm and hot gas in the outflow}

Does the nuclear outflow in NGC~2782 contain 
cold molecular gas in addition to  hot and warm 
ionised gas? Fig. 3a  (Plate L3) shows two CO spurs 
(labelled O1 and O2) which are elongated along the 
kinematic minor axis, and  appear to originate from 
the center of the CO bar  where 
the starburst intensity peaks. 
The  CO isovelocity contours 
show clear kinks at the base of these spurs, indicating 
deviations from circular motions (see Paper II). 
The spatial velocity plot (Fig. 5 of Paper II) 
shows non-circular motions of $\sim$ 30 km s$^{-1}$ 
in the two CO spurs.  
With the near side of the disk being north, 
these velocities are consistent with 
vertical outflow of gas out of the galaxy plane 
and/or radial inflows of gas in the galaxy plane. 
The fact that the spurs are elongated along the 
minor axis and lie inside the RC outflow bubbles 
(Fig. 3a, Plate L3) strongly suggests that the spurs 
contain outflowing gas. 

In order to determine the power needed to drive 
the outflow, we estimate the energy associated 
with the different phases of the ISM. 
With a Milky Way value for 
the CO-H$_{\rm 2}$ conversion factor, 
a radial vertical outflow velocity of  
(30 km s$^{-1}$/Sin $i$ ),  an inclination $i$ 
of 30 $\deg$ (see paper II), the CO spurs have 
$\sim$  $ 2 \times 10^7$ $M_{\sun}$ of  $H_{\rm 2} $ 
and a kinetic energy of $\sim$ $8 \times 10^{53}$  erg. 
The warm  ionised gas in the optical nebula has a mass 
of  $\sim$ $1.2 \times 10^{4} M_{\sun}$, a kinetic energy 
of $\sim$ $7 \times 10^{51}$ erg (Boer et al. 1992), 
and a thermal pressure of 
$\sim$ $3 \times 10^{-9}$ dynes cm$^{-2}$ 
for n$_e$ $\sim$ 600  cm$^{-3}$ (see $\S$ 4). 
Using the X-ray luminosity of  $5 \times 10^{40}$ erg s$^{-1}$
(from fluxes in Kinney et al. 1984 for a spectral index
of 0)  and following Nulsen, Stewart and Fabian (1984), 
we find that the hot gas has a total mass of 
 $\sim$  $6 \times 10^7$ f$^{0.5}$ $M_{\sun}$, 
a thermal energy of 
 $\sim$  $2 \times 10^{56}$ f$^{0.5}$ erg, 
and  a pressure of   
 $\sim$  $3 \times 10^{-10}$ f$^{-0.5}$ dynes cm$^{-2}$. 
We assumed a temperature
of $3 \times 10^{7}$ K, a radius of 15'' (2.5 kpc) 
and a volume filling factor f for the hot gas.
The above  estimates suggest that 
cold molecular gas and hot ionised gas 
dominate the mass and energy in the outflow of NGC~2782. 
Although the outflows in M82 and NGC~2782 appear to be 
in different dynamical phases, they contain 
similar masses of cold, warm, and hot gas. In M82, 
these masses are  estimated to be $\sim$ 
a few  $\times 10^7$ $M_{\sun}$ (Stark \& Carlson
1984), $ 2 \times 10^5$ $M_{\sun}$ (Heckman et al. 1990) and 
$ 1 \times 10^7$ $M_{\sun}$ (Schaaf et al. 1989) respectively. 

The optical linewidths and line ratios in the 
central region of NGC~2782 do not indicate 
the presence of an AGN (see $\S$3). 
Furthermore, NGC~2782 has a FIR luminosity 
($\sim$ 10$^{44}$ erg s$^{-1}$) and warm FIR 
colors (S$_{60\micron}$/S$_{100\micron}$~$\sim$~0.6) 
similar to those of  galaxies with the best evidence for 
supernova-driven winds (e.g., Lehnert \& Heckman 1996). 
(It is however noteworthy that the ratios of FIR to blue 
optical luminosity of both NGC~2782 and NGC~3079  are below 
1 rather than $\ge$~2 as is the case in most of these galaxies.)
We  investigate if star formation alone 
can power the outflow. 
We assume that the total energy of 
$\sim$ $2 \times 10^{56}$ f$^{0.5}$ erg in the outflow 
is a fraction  $x$ of the total energy generated by  
supernovae, where  $x$ depends on how efficiently 
the kinetic energy of supernovae is thermalised. 
With a mean energy of $\sim$ 2 $ \times 10^{51}$ 
erg per supernova (Mac Low \& McCray 1988) and an outflow 
timescale of  $4 \times 10^{6}$ yrs (Boer et al. 1992), 
the average supernova rate needed to power the outflow is 
$\approx$  (10$^{-2}$ f$^{0.5}$/$x$) yr$^{-1}$. 
Using the FIR luminosity of $2 \times 10^{10}$ $L_{\sun}$ 
as an estimate for the bolometric luminosity, and 
applying the starburst model of Rieke et al. (1980) 
gives a supernovae rate of $\approx$ 0.1 yr$^{-1}$.  
For a volume filling factor f of 0.1, the supernovae rate 
from the starburst 
can power the outflow if $x$ $\ge$ 5\%. Stellar winds can 
also be an important energy source in establishing a wind 
(Leitherer \& Heckman 1995).

NGC~2782 provides an unparalleled opportunity 
for setting observational constraints on 
the poorly-studied, early 
evolutionary phase of a starburst-driven outflow. 
Further observations of the dynamically 
young outflow in NGC~2782  can extend 
the study presented in this paper. 
In particular, future high resolution X-ray and near-infrared 
observations will help to further constrain the dynamical phases of 
both the southern and northern outflows ($\S$ 4), 
better estimate the poorly-known quantities  f and $x$, and 
study how the properties of the circumnuclear starburst 
influence the outflow it is driving.
\section{ Acknowledgements}
This work has been partially supported by an AAUWEF Fellowship, 
a Grant-in-Aid of Research from Sigma Xi (The  Scientific Research 
Society), an Amelia Earhart Fellowship, and an NSF grant AST-9322779. 
We thank D. J. Saikia and A. Pedlar for the 
5 GHz RC maps, J. Barnes at NOAO for the production of solitaires, 
and the High Energy Astrophysics Science Archive Research Center, 
provided by NASA's Goddard Space Flight Center, for the ROSAT X-ray map.
We thank  Richard Larson and referee Matthew Lehnert for helpful 
comments.
%
%
%
%

\newpage
\noindent
%
\bf
Fig. 1 (Plate L1)---  
\rm
The WIYN  R image of NGC~2782 with a  6.5$'$ (66 kpc) field of view 
reveals  the optical disk bracketed by two stellar tails. 
Except for three ripples at radii of 25$''$, 45$''$, and 60$''$, 
the optical disk is relatively undisturbed within a  radius 
of $\sim$ 1$'$ (10 kpc).

\medskip
\noindent

\medskip
\bf
Fig. 2 (Plate L2)--- 
\rm
(a) 
The  WIYN B-V image with a 1.6$'$ (16.3 kpc) field of view 
shows two dust lanes which are offset from the nucleus  
and extend  out to the inner ripple at r $\sim$ 25$''$ (4.2 kpc). 
Lighter regions in the image have redder B-V colors and darker 
regions have bluer B-V colors. 
The northern dust lane is redder and more prominent 
than the southern one. The dust lanes bracket the 
inner few kpc region which harbors  one of the most luminous 
starbursts among nearby (D$<$ 40 Mpc) spirals.
(b) 
The 0.8$''$ resolution  WIYN  H$\alpha$ image with 
a central 20$''$ (3.4 kpc) field of view, shows 
a central H$\alpha$ peak, an arc of star formation 
extending east-west, and a southern H$\alpha$ outflow 
which has  a well-defined bubble morphology with a  shell 
feature at its edge.

\medskip
\noindent
\bf
Fig. 3 (Plate L3)---
\normalsize
\bf
The starburst driven outflow: 
\rm
(a) 
5 GHz RC (contours, Saikia et al. 1994) 
on the $2.1'' \times 1.5''$ CO map (greyscale, from Paper II) 
for the central 20$''$ (3.4 kpc). 
It is likely that the main RC peak, as well as the 
secondary eastern and western peaks, are due to 
massive star formation. 
In contrast, the northern 
and southern RC bubbles which are elongated along the 
kinematic minor axis, are associated 
with little molecular gas and are 
probably part of the starburst-driven outflow.
The two CO spurs, labelled O1 and O2, 
are also elongated along the kinematic minor axis 
and lie inside the RC outflow bubbles.
The contour levels are 5, 10, 30, 50, 70, 90, 100 \% of the peak 
flux (3.97  mJy per beam). 
\rm
(b) 
\rm
The 0.8$''$ resolution  WIYN  H$\alpha$ image 
shows a region of star formation whose activity peaks between the 
two CO lobes, and a remarkable southern H$\alpha$ outflow bubble with 
an outer shell feature at $\sim$  6$''$ south. 
The cross marks the 5 GHz RC peak. 
The contour levels are 65, 75, 85, 90, 95, 100 \% of the peak 
flux. 
\rm
(c) H$\alpha$ (contours) on the the 5 GHz 
RC (greyscale). The H$\alpha$ and  RC peak 
are coincident within the uncertainties, and lie 
between the two CO lobes. 
The H$\alpha$ shell at  $\sim$ $6''$ (1  kpc) 
south  coincides with the shell in the southern 
RC bubble.
\rm
(d) The [O III] image (with 1.3$''$ seeing)  
has a similar bubble morphology as the outflowing H$\alpha$, 
and shows a strong  extranuclear peak at $\sim$  6 $''$ south. 
This extranuclear peak occurs in the region  where 
long-slit spectra shows a maximum in the intensity and 
FWHM of blueshifted [O III] lines.



\medskip
\noindent
\bf
Fig. 4 (Plate L4) --- 
\normalsize
(a) The H$\alpha$ image with a 76$''$ (13 kpc) field of view. 
This figure shows a string of HII regions which coincides with 
the first optical ripple at r $\sim$ 25$''$ (4.2 kpc), 
a bright northwestern H$\alpha$ arc, and 
diffuse H$\alpha$ loops and filaments which 
surround the central starburst region.  
(b) The ROSAT X-ray map (contour) of resolution $\sim$ 12$''$ 
on the H$\alpha$ image (greyscale). There 
is bright emission  in the central  8$''$ (1.4 kpc) radius 
where the optical and RC outflow bubbles lie, and 
fainter emission which extends further 
out. The faint X-ray emission is more extended  northwest
than southeast, and  it stretches out to the prominent 
northwestern  H$\alpha$ arc.

\end {document}